%% file: ms.tex
\documentclass[a4paper,conference]{IEEEtran}
\IEEEoverridecommandlockouts
\usepackage{cite}
\usepackage{amsmath,amssymb,amsfonts}
\usepackage{graphicx}
\usepackage{textcomp}
\usepackage{xcolor}
\def\BibTeX{{\rm B\kern-.05em{\sc i\kern-.025em b}\kern-.08em
    T\kern-.1667em\lower.7ex\hbox{E}\kern-.125emX}}

\usepackage{url}
\usepackage{hyperref}
\usepackage[retain-explicit-plus,binary-units]{siunitx}
\usepackage{csquotes}

\usepackage{fancyvrb}
\input{pygments.tex}

\input{preamble}

\usepackage{tikz}
\usetikzlibrary{decorations.pathreplacing, calligraphy, arrows.meta, positioning, shapes.multipart, backgrounds, fit, shapes, calc}

\usepackage{todonotes}

\def\sectionautorefname{Sect.}

\newcommand{\refline}[1]{Line~\ref*{#1}}
\newcommand{\reflines}[2]{\refline{#1}\,-\,\refline{#2}}

\def\CPP{C\raise.22ex\hbox{{\footnotesize +}}\raise.22ex\hbox{\footnotesize +}}

\begin{document}
\title{Accurate and Extensible Symbolic Execution of Binary Code based on Formal ISA Semantics%
\thanks{This work was supported by the German Federal Ministry of Education and Research (BMBF) within projects Scale4Edge under grant no.~16ME0127, ECXL under grant no.~01IW22002, and VE-HEP under grant no.~16KIS1342.}
}

\author{
\IEEEauthorblockN{Sören Tempel\IEEEauthorrefmark{1}, Tobias Brandt\IEEEauthorrefmark{2}, Christoph Lüth\IEEEauthorrefmark{3}\IEEEauthorrefmark{4}, Christian Dietrich\IEEEauthorrefmark{1} and Rolf Drechsler\IEEEauthorrefmark{3}\IEEEauthorrefmark{4}}%
\IEEEauthorblockA{\IEEEauthorrefmark{1}Institute of Operating Systems and Computer Networks, Technische Universität Braunschweig, Braunschweig, Germany}%
\IEEEauthorblockA{\IEEEauthorrefmark{2}Independent Researcher, Bremen, Germany}
\IEEEauthorblockA{\IEEEauthorrefmark{3}Cyber-Physical Systems, DFKI, Bremen, Germany}%
\IEEEauthorblockA{\IEEEauthorrefmark{4}Institute of Computer Science, University of Bremen, Bremen, Germany}
tempel@ibr.cs.tu-bs.de, tobbra91@gmail.com, christoph.lueth@dfki.de, dietrich@ibr.cs.tu-bs.de, drechsler@uni-bremen.de
\vspace{-1em}
}

\maketitle

\begin{abstract}
Symbolic execution is an \acs{SMT}-based software verification and testing technique.
Symbolic execution requires tracking performed computations during software simulation to reason about branches in the software under test.
The prevailing approach on symbolic execution of binary code tracks computations by transforming the code to be tested to an architecture-independent \ac{IR} and then symbolically executes this \ac{IR}.
However, the resulting \ac{IR} must be semantically equivalent to the binary code, making this process complex and error-prone.
The semantics of the binary code are specified by the targeted \ac{ISA}, commonly given in natural language and requiring a manual implementation of the transformation to an \ac{IR}.
In recent years, the use of formal languages to describe \ac{ISA} semantics in a machine-readable way has gained increased popularity.
We investigate the utilization of such formal semantics for symbolic execution of binary code, achieving an accurate representation of instruction semantics.
We present a prototype for the \mbox{RISC-V} \ac{ISA} and conduct a case study to demonstrate that it can be easily extended to additional instructions.
Furthermore, we perform an experimental comparison with prior work which resulted in the discovery of five previously unknown bugs in the \ac{ISA} implementation of the popular \ac{IR}-based symbolic executor \emph{angr}.
\vspace{-1em}
\end{abstract}

\input{body.tex}

\bibliographystyle{IEEEtran}
\bibliography{main}

\end{document}

%% file: preamble.tex
\usepackage{xspace}
\def\thing{\textsc{BinSym}\xspace}
\def\libriscv{\textsc{LibRISCV}\xspace}

\usepackage{acronym}

\acrodef{SE}{symbolic execution}
\acrodef{IR}{intermediate representation}
\acrodef{DSL}{domain-specific language}
\acrodef{SUT}{software-under-test}
\acrodef{SMT}{satisfiability modulo theories}
\acrodef{ISA}{instruction set architecture}
\acrodef{BDD}{binary decision diagram}

\def\eg{e.g.,\xspace}
\def\ie{i.e.,\xspace}

\usepackage{etoolbox}
\usepackage[textsize=tiny,textwidth=1.4cm]{todonotes}
\definecolorseries{test}{rgb}{grad}[rgb]{.95,.55,.55}{11,11,17}
\resetcolorseries[10]{test}
\newcommand{\addtodoeditor}[1]{%
  \colorlet{#1}{test!!+!50}
  \expandafter\def\csname#1\endcsname##1{%
	\todo[color=#1,]{\sffamily\textbf{\uppercase{#1}:} ##1}\xspace%
  }
  \expandafter\newcommand\csname#1i\endcsname [1]{%
		\todo[inline, color=#1]{\sffamily\textbf{\uppercase{#1}:} {##1}}\xspace%
  }
}
\forcsvlist\addtodoeditor{st,cd}

\def\paragraph#1{\textbf{#1}\hspace{1em}\ignorespaces}
\def\dn#1{\tikz[baseline]\node[anchor=base,circle,draw,fill=white,inner sep=1pt,minimum width=0em]{#1};}

%% file: body.tex
\section{Introduction}
\label{sec:introduction}

Program analysis and software testing has the goal to validate, or even verify, that certain properties hold on the different paths through a \ac{SUT}.
In comparison to unit testing, \ac{SE}~\cite{cadar2008klee} allows the exploration of more, or even all, paths by interpreting the \ac{SUT} in a symbolic domain.
Instead of concrete values ($X = 5$), we propagate symbolic values ($X > 5$) through the \ac{SUT}.
This requires us to bridge a \emph{semantic gap} and \emph{translate} the \ac{SUT} (\eg given as C code) to symbolic expressions that manipulate symbolic values.
Later on, the \ac{SE} engine can formally reason about branches in the \ac{SUT} by using a constraint solver to check the feasibility of symbolic branch conditions.

\begin{figure}[t]
  \centering
  \includegraphics[page=1,width=0.7\linewidth]{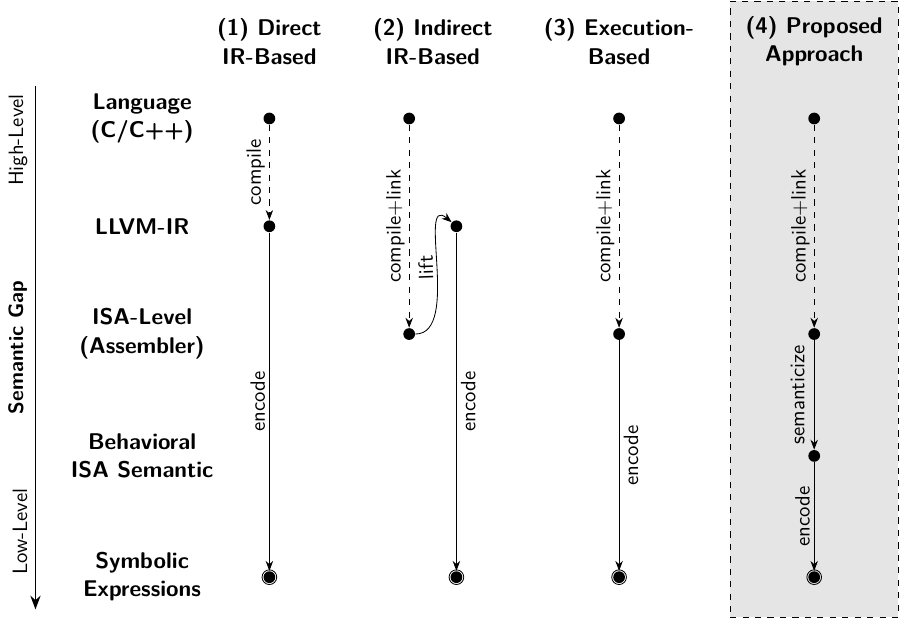}
  \caption{Different translation methodologies for symbolic execution.}
  \vspace{-1.5em}
  \label{fig:prior-work}
\end{figure}

While many translation methodologies (see \autoref{fig:prior-work}) were proposed~\cite{baldoni2018survey}, they all exhibit some shortcomings:
(1) \emph{Direct IR-based} translation, as employed by the popular KLEE~\cite{cadar2008klee} engine, first lowers the program to a compiler's \ac{IR} (\ie LLVM-IR) using the existing toolchain, before \emph{symbolizing} each \ac{IR} instruction to a symbolic expression.
While this approach captures compiler bugs, it is unable to handle inline assembly, compiler intrinsics, and third-party binary code.
(2) The \emph{Indirect IR-based} approach~\cite{brumley2011bap,djoudi2015binsec,shoshitaishvili2016angr} tackles this by lifting the final binary code back up to the IR level before symbolizing it down again to symbolic expressions.
This involves two semantically-rich translation steps:
First, the lifter has to capture the \ac{ISA} semantic, which is often only available as thousands of pages of informal, natural-language specification (\eg ARM manual~\cite{armv8a-arm}), making lifter construction an erroneous endeavor~\cite{dasgupta2020validation,soomin2017irtesting}.
Second, compiler \acp{IR} are usually implementation-defined and, even if formal semantics exists~\cite{zakowski:21:icfp}, then only as a secondary artifact that is prone to deviation.
(3) The \emph{Execution-based} method~\cite{tempel2023libriscv,yun2018qsym,herdt2019rvx,currie2000dsp} avoids this error-prone detour by weaving the \ac{SE} engine into execution(s) of the \ac{SUT}.
With an \ac{ISA}-level interpreter or by trap-stepping the actual machine, symbolic expressions are emitted on the fly.
While this avoids the \ac{IR} problem, capturing the instruction semantics \emph{correctly} and \emph{completely} is still challenging, especially when confronted with feature-rich\footnote{The ARMv8 Base \ac{ISA} alone has 472 instructions.} or rapidly-evolving extensible \acp{ISA}.%
\unskip\kern-1pt\footnote{RISC-V has 41 ratified extensions, 12 of them newly ratified in 2024~\cite{riscv:extensions}.}

\begin{figure*}
	\centering
        \includegraphics[page=1,width=0.8\linewidth]{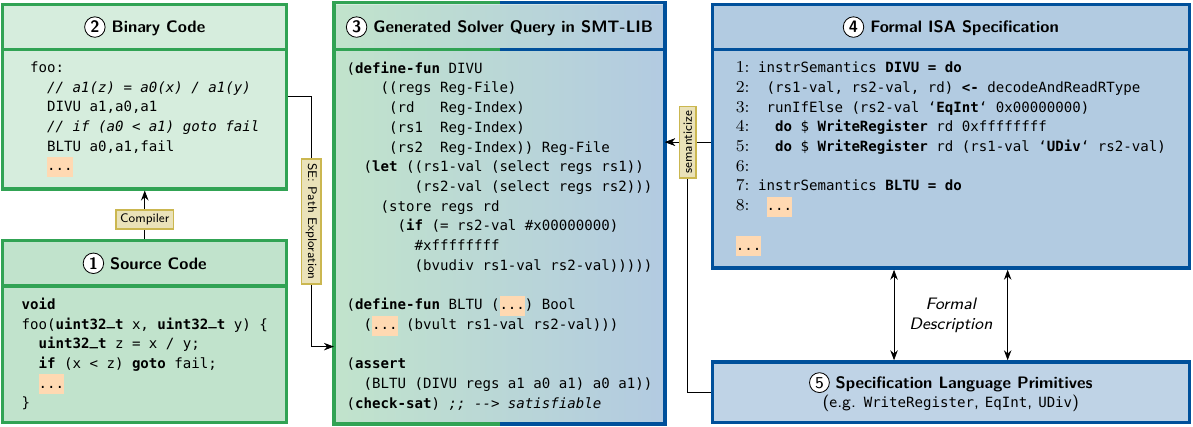}
	\caption{Generation of an SMT solver query for an exemplary branch condition in a SUT (in binary form) based on a formal ISA specification.}
	\vspace{-1em}
	\label{fig:approach}
\end{figure*}

\textbf{About this paper:}
Given the benefits and shortcomings of existing methods, we argue that (a) \ac{SE} engines should represent the program with a formal \enquote{intermediate representation} but (b) that it should be located \enquote{below} the \ac{ISA} level and explicitly designed to capture \ac{ISA} semantics (see \autoref{fig:approach}).
This would give us build-chain coverage and the possibility to independently verify the \ac{ISA}-level \enquote{semantification} step while retaining a strict lowering hierarchy.

Fortunately, \emph{formal \ac{ISA} specifications}~\cite{armstrong2019sail,bourgeat2022flexible,tempel2023libriscv} emerged in recent years to tackle the problem of \ac{ISA} manuals with thousands of pages of natural-language specification.
In a nutshell, these projects provide a formal, machine-readable \ac{DSL} to express \ac{ISA}-level semantics and come with tooling to work with these specifications (\eg derive theorem-prover definitions).
This direction seems so promising that \mbox{RISC-V}~\cite{riscv2019formal} and ARM~\cite{reid2016a} have recognized their potential and provide official formal specifications.

For design automation, utilization of formal \ac{ISA} specifications is advantageous as---once formally specified---a variety of tools can be derived from the specification, reducing development time and thereby the time-to-market.
For example, we can combine them with concurrency models~\cite{armstrong2021}, perform fault-injection~\cite{dietrich2022sailfail}, verify binary code~\cite{sammler2022}, prove security properties~\cite{bauereiss2022verified}, and derive emulators and documentation~\cite{armstrong2019sail}.

We argue that formal \ac{ISA} specifications and their \acp{DSL} are perfectly suited to make \ac{SE} extensible, accurate, and (potentially) easier to verify.
With this paper, we claim the following contributions:
\begin{itemize}
\item We present \mbox{\thing}, an \ac{SE} engine for \mbox{RISC-V} binary code that facilitates an executable formal specification.
\item We conduct a case study to demonstrate the extensibility of our approach with custom instruction set extensions.
\item We validate the accuracy of \mbox{\thing} by an experimental comparison to existing \ac{SE} engines, leading to the discovery of five previously unknown lifter bugs in prior work.
\end{itemize}

\section{Formal ISA Semantics for Symbolic Execution}
\label{sec:approach}

We give a short primer on \acl{SE} and discuss the benefits of formal \ac{ISA} semantics for the \ac{SE} of binary code.

\subsection{Background on Symbolic Execution}

As \ac{SE} is a simulation-based software verification technique, it executes the \ac{SUT} to explore its state space by enumerating reachable execution paths.
Unlike unit testing, which runs the program with concrete values, \ac{SE} replaces selected concrete inputs with symbolic ones that represent a set of continuously constrained concrete values.
The \ac{SE} engine manages this process by tracking constraints and iteratively selecting the next path to explore, a process referred to as \emph{path selection}~\cite[\sectionautorefname~2.2]{baldoni2018survey}.
During execution of a selected path, the \ac{SE} engine collects constraints on symbolic values up to a previously unvisited branch.
The constraints are obtained by \emph{translating} the performed operations on symbolic values to symbolic expressions, and a constraint solver checks which upcoming branch targets are feasible under the collected constraints.
Usually, we use the \ac{SMT} formalism (\ie \ac{SMT} bitvector theory) to perform constraint checking.

In a nutshell, the classical \ac{SE} engine \emph{repeatedly} executes program paths with symbolic values on top of an \ac{SMT} solver.
More information on \ac{SE} is available in the survey~\cite{baldoni2018survey}.

To illustrate binary-based \ac{SE}, we compile an example C program \dn{1} in \autoref{fig:approach} to binary code \dn{2}.
The C function \texttt{foo()} performs division of two 32-bit unsigned integers and ensures that the result is greater than or equal to the dividend (failing otherwise).
Since the control flow forks at the \texttt{BLTU} branch instruction, the engine selects the path \texttt{[DIVU, BLTU]} and translates it, including the arguments, to the \ac{SMT} expression \dn{3}, which we show in the standard \mbox{SMT-LIB}~\cite{smtlib}  format.
The engine then invokes an \ac{SMT} solver to check (\texttt{check-sat}) if the expression is satifisfiable resp. if it is possible to take the branch.
Conceptually, an \ac{SE} engine repeatedly generates such queries for each encountered branch in the \ac{SUT}.
The example demonstrates the significant semantic gap between \ac{ISA} instructions \dn{2} and \ac{SMT} expressions \dn{3} which the \ac{SE} engine has to bridge -- a translation that is known to be error-prone if done with handwritten translators~\cite{soomin2017irtesting,dasgupta2020validation}.

\subsection{ISA Semantics and SMT Translation}

The core idea of our proposed \ac{SE} approach is: formal \ac{ISA} semantics, which
capture the behavior of instructions in a machine-readable specification language, are well suited as an intermediate step between binary code and \ac{SMT} expressions.
In our running example (\autoref{fig:approach}), the normative \ac{ISA} semantics are provided in a machine-readable formal language \dn{4}, which expresses instruction behavior using a \ac{DSL} \dn{5} (further explained in \autoref{sec:libriscv}).
Different \acp{DSL} have been presented in prior work to express such semantics~\cite{reid2016a,armstrong2019sail,selfridge2019grift,tempel2023libriscv}.

By taking the specific \ac{ISA} semantics into account, an interesting edge case in the \ac{SUT} becomes apparent that has to be accounted for:
in the C programming language, division by zero is undefined behavior, and the compiler can assume that it never happens. 
Therefore, no check for \(y \equiv 0\) is inserted before the unsigned division instruction \texttt{DIVU} by the compiler.
However, the formal \texttt{DIVU} semantics \dn{4} show that the unsigned division of our \ac{ISA}---instead of trapping---returns a value with all bits set if the divisor is zero.
In this case, $z$ becomes \texttt{0xffffffff}, which is larger than most 32-bit $x$ values, whereby the \texttt{fail} branch is actually reachable with \(y = 0\).
This is contrary to the intuition of programmers reading \dn{1} as division usually makes numbers smaller, not larger.
This emphasizes the need for a binary-based, \ac{ISA}-specific \ac{SE}.

To uncover such problems, the \ac{SE} engine has to operate on the binary-level and translate the binary code instruction semantics to an equivalent \ac{SMT} representation.
As our example suggests, the translation from \dn{2} to \dn{3} is not straightforward as it requires us to capture %
(a) arithmetic and logic operations as well as %
(b) interactions with the hardware state (\eg the register file).
Note also how the \texttt{DIVU} representation in \dn{2} advances the system state by returning a new register file, in \autoref{fig:approach} we omit details for clarity of exposition (e.g. program counter handling).
The emitted \ac{SMT} must be accurate wrt. the \ac{ISA} specification, otherwise \ac{SUT} edge cases may be missed.

Instead of directly jumping from \ac{ISA} to \ac{SMT}, our approach uses the primitives of a formal specification as an abstraction layer.
That is, instead of directly translating the \texttt{DIVU} instruction to \ac{SMT}, we employ divide-and-conquer and translate the individual language primitives \dn{5} in which the semantics are expressed.
For example, the \texttt{WriteRegister} primitive is translated as a write to an \ac{SMT} array (\texttt{store}).
By building on these primitives, \ac{SE} engines also become more extensible: as long as new instructions can be expressed in terms of existing specification primitives, our \ac{SE} approach can be easily extended.
In \autoref{sec:case-study}, we will demonstrate this capability.

Our example illustrates the benefits of binary-code analysis and affirms that even supposedly simple code is expanded to complex \ac{SMT} representations.
Formal specifications of \ac{ISA} semantics help us to mitigate this complexity, reducing the potential for errors in the translation and enabling extensibility.

\section{Application to RISC-V Binary Code}
\label{sec:formalse}

In this section, we present \mbox{\thing}, a prototype implementation of our proposed approach that symbolically executes 32-bit binary code for the open standard \mbox{RISC-V}~\cite{riscvunpriv} architecture.
We chose \mbox{RISC-V} for our prototype as, due to its openness, it has enabled a large body of research on formal \ac{ISA} specifications and even provides an official golden formal specification~\cite{riscv2019formal}.
Furthermore, \mbox{RISC-V} is a modular architecture, \ie it consists of a base instruction set and optional extensions, which are implemented on top and can be combined as needed.
Therefore, it benefits immensely from an extensible \ac{SE} approach as the specification is constantly expanding, requiring binary analysis tools to \enquote{catch up} to it.

\subsection{Executable Formal Specifications}
\label{sec:libriscv}

A variety of different formal \ac{ISA} specifications for \mbox{RISC-V} have emerged in recent years which target different use cases~\cite{selfridge2019grift,armstrong2019sail,bourgeat2022flexible,tempel2023libriscv}.
Since software execution is the focus of our work, we make use of an \emph{executable formal specification}.
Such specifications allow for the creation of custom \emph{modular interpreters}~\cite{liang1995modular} which are responsible for interpreting the language primitives used by the specification.
These interpreters are modular in the sense that the executable specification provides generic versions of essential components (\eg the register file or memory) which can be reused by the interpreter.
Different executable specifications have been presented in prior work~\cite{selfridge2019grift,bourgeat2022flexible,tempel2023libriscv}.
For our \mbox{\thing} prototype implementation, we build upon the open source \mbox{\textsc{LibRISCV}}~\cite{tempel2023libriscv} specification. 
Like many executable formal specifications~\cite{bourgeat2022flexible,selfridge2019grift}, \mbox{\textsc{LibRISCV}} describes the \ac{ISA} using a \ac{DSL} that is embedded into the functional and strongly-typed general-purpose programming language Haskell.

In order to illustrate this \ac{DSL}, we discuss the formal description of the \mbox{RISC-V} \texttt{DIVU} instruction as provided by \mbox{\textsc{LibRISCV}}.
We have already seen this formal description in top-right corner \dn{4} of \autoref{fig:approach} and in the following, we describe it in greater detail.
The formal \texttt{DIVU} description in \autoref{fig:approach} starts off by specifying the operands of the instruction in Line~2.
As mandated by the \mbox{RISC-V} specification, \texttt{DIVU} is an \mbox{R-Type} instruction with three register operands (\texttt{rs1}, \texttt{rs2}, and \texttt{rd}).
The instruction semantics are described in terms of these operands (Line~3\,-\,Line~5).
If the divisor (\texttt{rs2}) is zero (Line~3), then the destination register (\texttt{rd}) has all bits set (Line~4).
Otherwise, standard unsigned division is performed on the dividend (\texttt{rs1}) in Line~5.
This description is entirely abstract and does \emph{not assume} a specific representation of operands.

\subsection{BinSym: A Modular Symbolic Interpreter}

The \texttt{DIVU} example serves to illustrate that \mbox{\textsc{LibRISCV}} abstractly describes instruction semantics in terms of several language primitives: (1) stateful ones and (2) arithmetic/logic primitives.
Since \mbox{\textsc{LibRISCV}} is an executable formal specification, it takes \mbox{RISC-V} binary code (in the ELF format) as an input and converts the instruction stream to a continuous, lazily evaluated sequence of these primitives.
The aforementioned modular interpreter is then responsible for processing this sequence.
As an example, prior work has presented a concrete interpreter and an interpreter performing dynamic information flow tracking~\cite{tempel2023libriscv}.
For \mbox{\thing}, we implemented a symbolic modular interpreter that utilizes the language primitives as an abstraction layer for the implementation of a binary-level \ac{SE} engine.
In \autoref{fig:prior-work}, \mbox{\thing} implements the \emph{semanticize} step.

For \ac{SE}, instruction operands are symbolic values (\eg the value of the register \(x1\) may be symbolic).
As such, we need variants of the register file and memory that are capable of operating on such symbolic values.
Due to the utilization of an executable formal model, we were able to reuse existing components from \mbox{\textsc{LibRISCV}} for this purpose, such as a generic implementation of a register file which is parameterized over a value type.
This significantly reduced the effort required for our \mbox{\thing} prototype implementation and is one major benefit of an executable formal specification.
The symbolic representation of the hardware state enables us to symbolically interpret stateful \mbox{\textsc{LibRISCV}} language primitives such as \texttt{WriteRegister}.
Additionally, we had to map arithmetic and logic primitives to operations of an \ac{SMT} solver.
We make use of the Z3~\cite{moura2008z3} \ac{SMT} solver in \mbox{\thing}, translating arithmetic and logic operations of \mbox{\textsc{LibRISCV}} to Z3 \ac{SMT} solver operations.
This is the \emph{encode} step from \autoref{fig:prior-work}.

The outlined translation enables us to propagate and track symbolic values throughout program execution.
Based on the propagated values, we can symbolically reason about branch points during execution of the \ac{SUT}.
That is, every time we encounter a branch (denoted via the \texttt{runIfElse} language primitive) that depends on a symbolic value, we can use Z3 to check if both the true and the false case are satisfiable and if so, explore both in parallel.
For example, if a \ac{SUT} executes a \mbox{RISC-V} \texttt{DIVU} instruction with a symbolic divisor operand, we construct an \ac{SMT} query to check if it is possible for the divisor to be zero/non-zero under the current constraints.
On the technical side, \mbox{\thing} implements a so-called offline executor, which continuously restarts execution of the \ac{SUT} with input values obtained for branch points from the solver~\cite[\sectionautorefname~2.4]{baldoni2018survey}.
Specifically, it implements dynamic \acl{SE}~\cite[\sectionautorefname~2.1]{baldoni2018survey} with depth-first search path selection~\cite[\sectionautorefname~2.2]{baldoni2018survey} and address concretization~\cite[\sectionautorefname~3.2]{baldoni2018survey}.
These are well-established \ac{SE} algorithms (details in the references).

Our \mbox{\thing} prototype implements the entire \ac{SE} engine for \mbox{RISC-V} binary code in only 1000~LOC in Haskell with 1500~LOC of \libriscv specification.
This reduced complexity can be attributed to the utilization of an executable formal specification and its implementation as a modular interpreter.
Please note that while our prototype focuses on \mbox{RISC-V}, it is by no means limited to this architecture.
Formal semantics are available for many \acp{ISA} and prior work has demonstrated that it is feasible to capture the semantics of different \acp{ISA} using a common set of language primitives~\cite{armstrong2019sail}.

\section{Case Study: Supporting A Custom Instruction}
\label{sec:case-study}

With the advent of \mbox{RISC-V}, custom instructions are becoming increasingly popular for power savings in embedded systems or increased throughput in high-performance processors~\cite[\sectionautorefname~2]{cui2023riscvext}.
Utilization of formal \ac{ISA} semantics significantly eases supporting---and experimenting with---custom instructions during design space exploration.
Once formally described, documentation, simulators, fault-injection tooling, et cetera can be derived~\cite{dietrich2022sailfail,armstrong2019sail,tempel2023codegen}.
Thereby, formal semantics significantly reduce the effort required to support custom instructions throughout the development process, contributing to a reduced time-to-market.
In the following, we conduct a case study with an exemplary custom instruction to demonstrate that---once formally specified---custom instructions can be easily supported in our proposed \ac{SE} approach.

\begin{figure}
	\vspace{-1em}
	\input{code/instr_dict.tex}
	\caption{YAML \texttt{riscv-opcodes} description of a custom \texttt{MADD} instruction.}
	\vspace{-1em}
	\label{figure:madd-decode}
\end{figure}

For our case study, we define a new non-standard \texttt{MADD} instruction which takes three register operands (\texttt{rs1}, \texttt{rs2}, \texttt{rs3}) and combines multiplication and addition into a single instruction computing: \((rs1 \times rs2) + rs3\).
In order to support this custom instruction in our \ac{SE} engine, we first need to specify how it is encoded.
For this purpose, \mbox{\textsc{LibRISCV}} utilizes the existing \texttt{riscv-opcodes}\footnote{\url{https://github.com/riscv/riscv-opcodes}} instruction format provided by the \mbox{RISC-V} Foundation.
\autoref{figure:madd-decode} provides the description of the \texttt{MADD} encoding in this format.
Essentially, the format defines two bitmasks (\refline{line:opcode_mask} and \refline{line:opcode_match}) which can be used to uniquely identify the instruction's opcode.
Further, it specifies which well-known instruction operand fields are used by the instruction (\refline{line:opcode_field}).
The existing \mbox{\textsc{LibRISCV}} tooling automatically generates the decoding code from this description~\cite[\sectionautorefname~4.1]{tempel2023libriscv}.

In addition to the instruction encoding, we also need to specify the instruction semantics in terms of the language primitives provided by \mbox{\textsc{LibRISCV}}.
The formal description of our custom \texttt{MADD} instruction is shown in \autoref{figure:madd-desc}.
The instruction is decoded as an R4-Type instruction (\refline{line:madd_decode}), then sign-extension and multiplication of \texttt{rs1} and \texttt{rs2} is performed (\reflines{line:madd_mul_start}{line:madd_mul_end}), the lower 32-bit are incremented by \texttt{rs3} and stored in the destination register (\refline{line:madd_add}).
The semantics of \texttt{MADD} can be expressed entirely in terms of existing \mbox{\textsc{LibRISCV}} language primitives.
As our \mbox{\thing} prototype already maps all of these primitives to \ac{SMT} semantics, no modifications of \mbox{\thing} are needed in order to support this instruction.
In total, we only had to integrate the 7 lines of YAML from \autoref{figure:madd-decode} and the 7 lines of Haskell code from \autoref{figure:madd-desc} into the formal \ac{ISA} specification to support \ac{SE} with \texttt{MADD}. 

\begin{figure}[b]
	\vspace{-1em}
	\input{code/madd.tex}
	\caption{Description of the custom \texttt{MADD} semantics in \mbox{\textsc{LibRISCV}}.}
	\label{figure:madd-desc}
\end{figure}

Naturally, the formal description of the \texttt{MADD} instruction (as shown in \autoref{figure:madd-desc}) can not only be used for symbolic semantics but also for other design automation tooling.
This serves to illustrate that our approach can be easily extended to support additional instructions and, moreover, integrates well with a design flow centered around formal \ac{ISA} semantics.

\section{Evaluation}
\label{sec:performance}

We evaluate our approach on five programs: three real-world modules from the RIOT operating system~\cite{baccelli2018riot} (\texttt{base64-encode}, \texttt{clif-parser} and \texttt{uri-parser}) and two synthetic benchmark applications (\texttt{bubble-sort} and \texttt{insertion-sort}).
The latter benchmark applications have also been used in prior work for evaluation purposes~\cite[\sectionautorefname~4.2]{corteggiani2018inception}.
All programs have been compiled for the \mbox{32-bit} \mbox{RISC-V} architecture; we can therefore only compare against prior \ac{SE} work that supports this architecture.
We are presently aware of the following \ac{SE} engines that fulfill this criterion: angr~\cite{shoshitaishvili2016angr}, \mbox{\textsc{BinSec}}~\cite{djoudi2015binsec}, and \mbox{\textsc{SymEx-VP}}~\cite{tempel2022symex}.

We are interested in two evaluation aspects: (a) does our work discover the same amount of execution paths as prior work and (b) does our work achieve competitive \ac{SE} performance?
Therefore, we symbolically executed the programs with a fixed-size input of symbolic values.
All benchmarks are fully explorable, \ie all engines can discover the same amount of execution paths on all benchmarks.
Additionally, all tested \ac{SE} engines have been configured to use the same version of Z3~\cite{moura2008z3} to avoid benchmarking the solver.
We focus on path coverage here, not the detection of bugs in the \ac{SUT}.

\subsection{Exploration Results}

\begin{table}
	\caption{Amount of execution paths found by different SE engines.}
	\centering
	\scriptsize
	\input{tables/paths-found.tex}
	\label{table:paths-found}
	\vspace{-1em}
\end{table}

We compare the results of the symbolic exploration in \autoref{table:paths-found}.
This table contains one row for each tested program and compares the execution paths found by the aforementioned \ac{SE} engines.
As evident by the gathered data, angr fails to discover all execution paths in real-world modules from the RIOT operating system.
Specifically, angr misses 6125~paths in the \texttt{base64-encode} program and 46~paths in the \texttt{uri-parser} program (see the cells marked with a \textdagger{} in \autoref{table:paths-found}).
These execution paths were found by all other tested engines, including our \mbox{\thing} prototype.
Further debugging of angr revealed that these execution paths are missed due to implementation errors in the \mbox{RISC-V} lifter provided by angr.
In total, \emph{we found five previously unknown bugs} which have been reported, acknowledged, and fixed by angr developers:\footnote{\url{https://github.com/angr/angr-platforms/pull/64}}
\begin{enumerate}
	\item The arithmetic shift operation (\eg as used in the \texttt{SRA} instruction) was modeled incorrectly in the lifter.
	\item The R-Type shift instructions used the lower bits of the register index, not the register value, as a shift amount.
	\item The lifted load instructions did not correctly zero- and sign-extend the resulting register value.
	\item The shift amount for I-type shift instructions was treated as a signed, instead of unsigned, integer value.
	\item The signed comparison instructions compared for unsigned instead of signed integer equality.
\end{enumerate}
All inaccuracies are caused by programming errors in the manual implementation of the natural language \ac{ISA} specification within the lifter provided by angr.
These programming errors can result in both false-positives and false-negatives during automated \ac{SE}-based software testing.
As an example, consider the C code in \autoref{fig:angr-bug} which causes \ac{SUT} analysis using angr to result in both a false-negative and false-positive.
This code calculates a bitmask based on a function parameter \texttt{x}, if \mbox{\texttt{x == 1}} it expects the bitmask to be \texttt{0x80000000} and ensures that this is the case using an assertion (\refline{line:false_positive}).
The code is compiled to use an I-type shift instruction, for such instructions angr incorrectly treats the 5-bit shift amount immediate as a signed integer in two's complement.
Therefore, it ends up shifting by \texttt{-1} instead of \texttt{31}.
This results in a spurious assertion failure as the bitmask is not equal to \texttt{0x80000000} if \mbox{\texttt{x == 1}} (\ie angr finds a false-positive in the \ac{SUT}).
Similarly, the code contains an additional assertion which (incorrectly) assumes that \mbox{\texttt{x == 1}} is the only input to generate the bitmask value \texttt{0x80000000}.
However, this assertion is incorrect as there are other values for \texttt{x} which result in such a bitmask value (\eg \texttt{0xffffffff}).
Unfortunately, due to the issue outlined above, angr fails to find such an input, resulting in a false-negative.
In real-world code, large amounts of false-positives and false-negatives can significantly hinder adaption of \ac{SE}~\cite{heckman:2011:insof} and must therefore be avoided.
With our approach, we can---conceptually---avoid errors in the implementation of the \ac{ISA} semantics by building on a normative and therefore authoritative formal \ac{ISA} specification.

\begin{figure}
	\centering
	\input{code/angr_bug.tex}
	\caption{Example code which results in false-positive and false-negative when analyzed using the existing angr \ac{SE} engine due to a bug in the RISC-V lifter.}
	\vspace{-1em}
	\label{fig:angr-bug}
\end{figure}

\subsection{Performance Comparison}

\begin{figure}
	\centering
	\includegraphics[width=\linewidth]{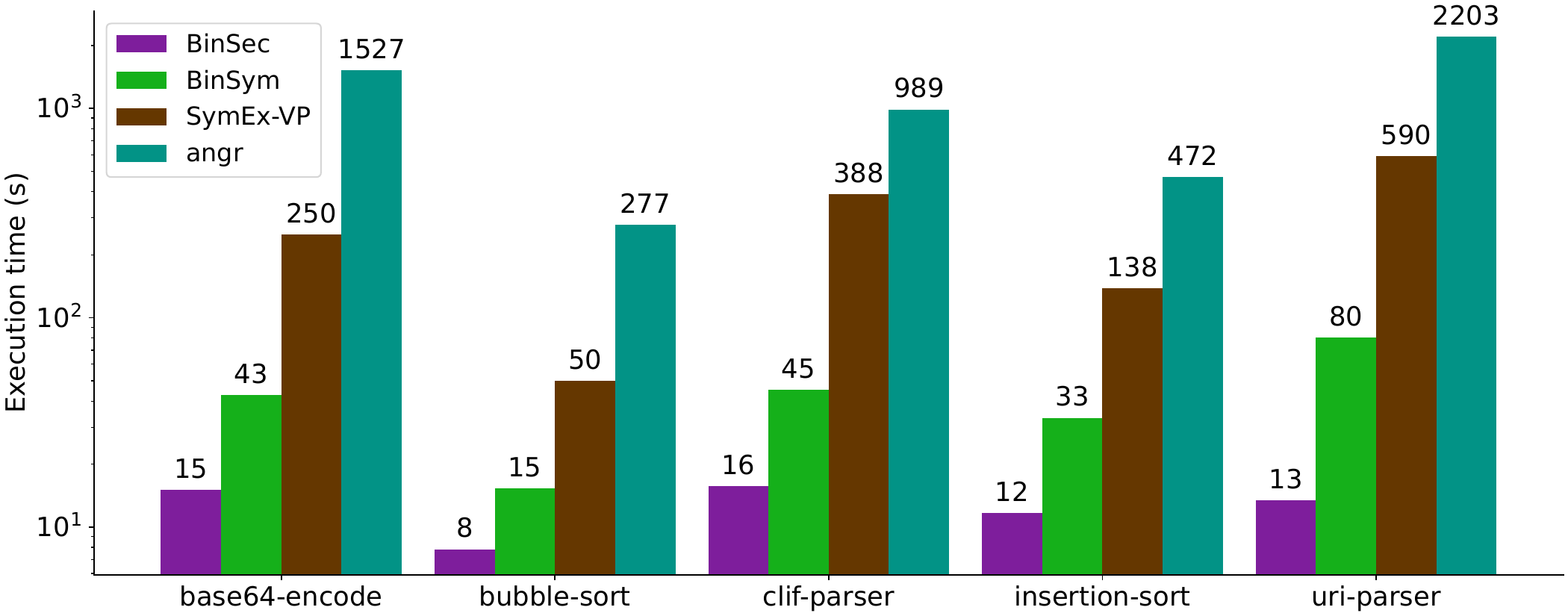}

	\caption{Total execution time as an arithmetic mean over five executions per benchmark. The y-axis is logarithmic as results for angr constitute an outlier.}
	\vspace{-1em}
	\label{fig:speed-plot}
\end{figure}

Utilizing the fixed version of angr, we experimentally compare \ac{SE} performance.
To this end, we executed all programs five times with each engine on an Intel Xeon Gold 6240 Linux system.
The arithmetic mean over all five executions is visualized as a grouped bar chart in \autoref{fig:speed-plot}.
In this figure, the absolute execution time---as the arithmetic mean over all five executions---is given logarithmically in seconds on the \mbox{y-axis} while the \mbox{x-axis} lists the benchmarks.
For each benchmark, four bar charts are given which correspond to the aforementioned \ac{SE} engines; from left to right: \mbox{\textsc{BinSec}} (purple), \mbox{\thing} (green), \mbox{\textsc{SymEx-VP}} (brown) and angr (teal).
Maximum standard deviation across all executions is \SI{5}{\percent}, which we consider negligible.
The results can be reproduced using the provided evaluation artifacts~\cite{artifacts} and are consistent across all benchmark applications: \mbox{\textsc{BinSec}} is the fastest engine, followed by our \mbox{\thing} prototype implementation and prior work on \mbox{\textsc{SymEx-VP}}.
The slowest engine across all benchmarks is angr.
This is expected and congruent with prior work which saw similar results for angr and attributes its \enquote{lower execution rate to the fact that its symbolic reasoning is implemented in Python}~\cite[\sectionautorefname~5.4]{poeplau2019symcompare}.
Similarly, \mbox{\textsc{SymEx-VP}} executes software in a SystemC~\cite{systemc} simulation environment, which enables it to support interactions with hardware peripherals modeled in SystemC~\cite{systemc} but incurs a performance penalty~\cite[\sectionautorefname~3.2]{tempel2022symex}.
\mbox{\textsc{BinSec}} is not subject to these limitations and is one of the most mature and optimized \ac{SE} engines for binary code.
Hence, it achieves better simulation performance than \mbox{\thing}, which is a prototype of our translation approach, and subsequently lacks advanced optimizations in the exploration and solver components.

Nonetheless, overall, the results from \autoref{fig:speed-plot} show that \mbox{\thing} offers competitive execution time performance.
That is, the results indicate that the technologies and techniques used in \mbox{\thing} (\ie executable formal \ac{ISA} semantics) do not negatively impact performance in a way that utilization for \ac{SE} becomes infeasible.
Unfortunately, as pointed out by prior work, it remains challenging to isolate which design decisions contribute to overall \ac{SE} performance as part of an empirical comparison~\cite{poeplau2019symcompare}.
Since we now know that the overall performance is competitive, we plan to expand on the evaluation in future work by specifically investigating the impact of formal \ac{ISA} semantics on \ac{SMT} query complexity.

\section{Related Work}
\label{sec:related-work}

\textbf{Symbolic Execution} is an active research area on which Baldoni et~al.~\cite{baldoni2018survey} give a recent overview.
As \autoref{sec:introduction} already categorizes different translation methodologies to \ac{SMT}, we only discuss the further \ac{SE}-related work here.
In contrast to \emph{source-based} \ac{SE}, of which KLEE~\cite{cadar2008klee} is most popular, \emph{binary-based} methods, like our \mbox{\thing}, do not require access to the source code.
While initial binary-based tools~\cite{chipounov2011s2e,davidson2013fie,corteggiani2018inception} built upon KLEE, and thus LLVM-IR, also other high-level \acp{IR}, such as Valgrind VEX~\cite{nethercote2007valgrind,shoshitaishvili2016angr} or DBA~\cite{bardin2011dba,djoudi2015binsec} were utilized.
These methods, which all rely on binary lifting~\cite{poeplau2019symcompare}, use the \ac{IR} as a semantically-rich \enquote{meta}-\ac{ISA} to unify the different \acp{ISA}.
We propose to use formal \ac{ISA} specifications, which were designed with verifiability in mind, to replace the lifting process with a straightforward per-instruction lowering.

\textbf{Translational Correctness} requires us to show the equivalence between binary code and symbolic expressions.
If we use a lifter in the process, we have to show two equivalences:
binary$\leftrightarrow$\ac{IR} and \ac{IR}$\leftrightarrow$symbolic expressions.
Previous validation efforts focused on the former~\cite{dasgupta2020validation,hendrix2019reopt,soomin2017irtesting} and revealed many bugs in existing tools.
This is in line with our findings (see \autoref{sec:performance}) that manual lifter design is an error-prone process.
With our proposed approach, the \ac{ISA} semantics come in the form of an independently test- and verifiable artifact to the process.
The aforementioned prior work also acknowledges that there are \enquote{few existing approaches to testing the correctness of binary lifters}~\cite{soomin2017irtesting}.
Unsurprisingly, there is even less work that goes beyond lifter correctness and also validates the symbolic semantics~\cite{correnson2023symcorrect,appel2011verismall,kuechel2022verified}.
Presently, such work operates on toy programming languages and, to the best of our knowledge, there is no prior work which concerns itself with the correctness of symbolic semantics for real-world binary code.
The formal representation of \ac{ISA} semantics in \mbox{\thing} is an ideal springboard to formally prove the equivalence between the \ac{ISA} semantics and the \ac{SMT} representation in future work.

\textbf{Formal Specifications} of programming semantics have already been used for \ac{SE} in prior work~\cite{lucanu2017framework}.
However, formal specifications of \ac{ISA} semantics have only gained increased relevance in recent years with Sail~\cite{armstrong2019sail} being the most extensive work and selected in 2019 as the official formal \mbox{RISC-V} specification~\cite{riscv2019formal}.
Utilization of such formal \ac{ISA} specifications for \ac{SE} of binary code is presently limited.
Goel et~al.~\cite{goel2014symex} use a partial \ac{ISA} model to perform automated proofs on x86 binaries through \ac{SE} based on \acp{BDD} instead of \ac{SMT}, limiting its applicability to general programs.
Prior work on retargetable tooling (\eg TSL~\cite{lim2013tsl} and CDT~\cite{ibing2015symex}) generates code for binary analysis tooling from an informal, non-executable architecture description.
Unfortunately, maintaining a diverse set of code generators for different analysis tasks is a laborious undertaking~\cite[\sectionautorefname~1]{bourgeat2022flexible}.
In contrast, our \mbox{\thing} prototype is based on an executable specification in the lineage of prior work on modular interpreters~\cite{liang1995modular}, enabling reuse of existing components and significantly reducing the \ac{SE} engine's complexity.

\section{Conclusion}

We present \mbox{\thing}, an approach for \acl{SE} of binary code that is accurate wrt. formal \ac{ISA} specifications.
Such specifications are easily extensible and increasingly adopted by modern \acp{ISA}.
The novelty of our approach is the utilization of an \emph{executable} formal specification: we derive the \ac{SE} engine as a modular interpreter for this specification, which reduces its complexity and helps to avoid bugs during the translation from \ac{ISA} instructions to \ac{SMT} expressions.

With the \mbox{\thing} prototype, we show that our approach allows to implement a complete \ac{SE} engine for RISC-V in 1000 lines of code, excluding the formal \ac{ISA} specification. 
We illustrate the extensibility with a custom instruction and experimentally compare \mbox{\thing} to three existing \ac{SE} engines.
Thereby, we discovered five previously unknown bugs in angr, a popular \ac{IR}-based symbolic executor, which can result in missing or incorrect \ac{SE} results.
These bugs originate from the manually-written binary-to-\ac{IR} lifter in angr, highlighting the importance of deriving all parts of the testing toolchain from a single authoritative \ac{ISA} specification.
Further, our experiments indicate that \mbox{\thing} achieves competitive \ac{SE} simulation performance in comparison to existing prior work.
To stimulate further research on binary-based SE with formal \ac{ISA} specifications, \mbox{\thing} is available as open source software.\footnote{\url{https://github.com/agra-uni-bremen/binsym}}

%% file: code/instr_dict.tex
\begin{Verbatim}[commandchars=\\\{\},codes={\catcode`\$=3\catcode`\^=7\catcode`\_=8\relax},numbers=left,numbersep=5pt,fontsize=\footnotesize,xleftmargin=10pt]
\PY{n+nt}{madd}\PY{p}{:}
\PY{+w}{  }\PY{n+nt}{encoding}\PY{p}{:}\PY{+w}{ }\PY{l+s}{\PYZsq{}}\PY{l+s}{\PYZhy{}\PYZhy{}\PYZhy{}\PYZhy{}\PYZhy{}01\PYZhy{}\PYZhy{}\PYZhy{}\PYZhy{}\PYZhy{}\PYZhy{}\PYZhy{}\PYZhy{}\PYZhy{}\PYZhy{}\PYZhy{}\PYZhy{}\PYZhy{}\PYZhy{}\PYZhy{}\PYZhy{}\PYZhy{}\PYZhy{}1000011}\PY{l+s}{\PYZsq{}}
\PY{+w}{  }\PY{n+nt}{extension}\PY{p}{:}\PY{+w}{ }\PY{p+pIndicator}{[}\PY{n+nv}{rv\PYZus{}zimadd}\PY{p+pIndicator}{]}
\PY{+w}{  }\PY{n+nt}{mask}\PY{p}{:}\PY{+w}{ }\PY{l+s}{\PYZsq{}}\PY{l+s}{0x600007f}\PY{l+s}{\PYZsq{}}\PY{esc}{\label{line:opcode_mask}}
\PY{+w}{  }\PY{n+nt}{match}\PY{p}{:}\PY{+w}{ }\PY{l+s}{\PYZsq{}}\PY{l+s}{0x2000043}\PY{l+s}{\PYZsq{}}\PY{esc}{\label{line:opcode_match}}
\PY{+w}{  }\PY{n+nt}{variable\PYZus{}fields}\PY{p}{:}\PY{+w}{ }\PY{p+pIndicator}{[}\PY{n+nv}{rd}\PY{p+pIndicator}{,}\PY{+w}{ }\PY{n+nv}{rs1}\PY{p+pIndicator}{,}\PY{+w}{ }\PY{n+nv}{rs2}\PY{p+pIndicator}{,}\PY{+w}{ }\PY{n+nv}{rs3}\PY{p+pIndicator}{]}\PY{esc}{\label{line:opcode_field}}
\end{Verbatim}

%% file: code/madd.tex
\begin{Verbatim}[commandchars=\\\{\},codes={\catcode`\$=3\catcode`\^=7\catcode`\_=8\relax},numbers=left,numbersep=5pt,fontsize=\footnotesize,xleftmargin=10pt]
\PY{n+nf}{instrSemantics}\PY{+w}{ }\PY{k+kt}{MADD}\PY{+w}{ }\PY{o+ow}{=}\PY{+w}{ }\PY{k+kr}{do}
\PY{+w}{  }\PY{p}{(}\PY{n}{rs1}\PY{p}{,}\PY{+w}{ }\PY{n}{rs2}\PY{p}{,}\PY{+w}{ }\PY{n}{rs3}\PY{p}{,}\PY{+w}{ }\PY{n}{rd}\PY{p}{)}\PY{+w}{ }\PY{o+ow}{\PYZlt{}\PYZhy{}}\PY{+w}{ }\PY{n}{decodeAndReadR4Type}\PY{esc}{\label{line:madd_decode}}
\PY{+w}{  }\PY{k+kr}{let}
\PY{+w}{    }\PY{n}{multResult}\PY{+w}{ }\PY{o+ow}{=}\PY{+w}{ }\PY{p}{(}\PY{n}{sext}\PY{+w}{ }\PY{n}{rs1}\PY{p}{)}\PY{+w}{ }\PY{p}{`}\PY{k+kt}{Mul}\PY{p}{`}\PY{+w}{ }\PY{p}{(}\PY{n}{sext}\PY{+w}{ }\PY{n}{rs2}\PY{p}{)}\PY{esc}{\label{line:madd_mul_start}}
\PY{+w}{    }\PY{n}{multTrunc}\PY{+w}{  }\PY{o+ow}{=}\PY{+w}{ }\PY{n}{extract32}\PY{+w}{ }\PY{l+m+mi}{0}\PY{+w}{ }\PY{n}{multRes}\PY{esc}{\label{line:madd_mul_end}}
\PY{+w}{  }\PY{k+kt}{WriteRegister}\PY{+w}{ }\PY{n}{rd}\PY{+w}{ }\PY{o}{\PYZdl{}}\PY{+w}{ }\PY{p}{(}\PY{n}{multTrunc}\PY{+w}{ }\PY{p}{`}\PY{k+kt}{Add}\PY{p}{`}\PY{+w}{ }\PY{n}{rs3}\PY{p}{)}\PY{esc}{\label{line:madd_add}}
\end{Verbatim}

%% file: tables/paths-found.tex
\renewcommand{\arraystretch}{1.25}
\setlength{\tabcolsep}{5pt}

\begin{tabular}{l|r|r|r|r}
	Benchmark               & angr               & \textsc{BinSec} & \textsc{SymEx-VP} & \thing          \\\hline
	\texttt{base64-encode}  & 125 \textdagger{}  & 6250            & 6250              & 6250            \\
	\texttt{bubble-sort}    & 720                & 720             & 720               & 720             \\
	\texttt{clif-parser}    & 11424              & 11424           & 11424             & 11424           \\
	\texttt{insertion-sort} & 5040               & 5040            & 5040              & 5040            \\
	\texttt{uri-parser}     & 8194 \textdagger{} & 8240            & 8240              & 8240            \\
\end{tabular}

%% file: code/angr_bug.tex
\begin{Verbatim}[commandchars=\\\{\},codes={\catcode`\$=3\catcode`\^=7\catcode`\_=8\relax},numbers=left,numbersep=5pt,fontsize=\footnotesize,xleftmargin=10pt]
\PY{k+kt}{void}\PY{+w}{ }\PY{n+nf}{parse\PYZus{}word}\PY{p}{(}\PY{k+kt}{uint32\PYZus{}t}\PY{+w}{ }\PY{n}{x}\PY{p}{)}\PY{+w}{ }\PY{p}{\PYZob{}}
\PY{+w}{  }\PY{k+kt}{uint32\PYZus{}t}\PY{+w}{ }\PY{n}{mask}\PY{+w}{ }\PY{o}{=}\PY{+w}{ }\PY{n}{x}\PY{+w}{ }\PY{o}{\PYZlt{}}\PY{o}{\PYZlt{}}\PY{+w}{ }\PY{l+m+mi}{31}\PY{p}{;}
\PY{+w}{  }\PY{k}{if}\PY{+w}{ }\PY{p}{(}\PY{n}{x}\PY{+w}{ }\PY{o}{=}\PY{o}{=}\PY{+w}{ }\PY{l+m+mi}{1}\PY{p}{)}
\PY{+w}{    }\PY{n}{assert}\PY{p}{(}\PY{n}{mask}\PY{+w}{ }\PY{o}{=}\PY{o}{=}\PY{+w}{ }\PY{l+m+mh}{0x80000000}\PY{p}{)}\PY{p}{;}\PY{+w}{ }\PY{c+c1}{// FALSE\PYZhy{}POSITIVE\label{line:false_positive}}
\PY{+w}{  }\PY{k}{else}
\PY{+w}{    }\PY{n}{assert}\PY{p}{(}\PY{n}{mask}\PY{+w}{ }\PY{o}{!}\PY{o}{=}\PY{+w}{ }\PY{l+m+mh}{0x80000000}\PY{p}{)}\PY{p}{;}\PY{+w}{ }\PY{c+c1}{// FALSE\PYZhy{}NEGATIVE\label{line:false_negative}}
\PY{p}{\PYZcb{}}
\end{Verbatim}